%% file: randtreegen_full.tex
\documentclass[11pt]{article}
\usepackage[OT4]{fontenc}
\usepackage{fullpage}
\usepackage{amsmath,amsthm,amsopn,amsfonts}
\usepackage{enumerate}
\usepackage{ifpdf}
\usepackage{graphicx}
\usepackage{dsfont}

\newtheorem{theorem}{Theorem}
\newtheorem{lemma}[theorem]{Lemma}

\newtheorem{definition}[theorem]{Definition}

\newtheorem{fact}[theorem]{Fact}

\newenvironment{Proof}{\noindent {\em Proof:}
 \noindent}{\hfill \rule{0.5em}{2ex} }

\newenvironment{Proof-sketch}{\noindent {\em Sketch of the proof:}
 \noindent}{\hfill \rule{0.5em}{2ex} }

\newcommand{\COMMENTED}[1]{}

\newcommand{\TO}{\textbf{$\widetilde{O}$}}
\newcommand{\TX}{\textbf{$\widetilde{X}$}}
\newcommand{\HX}{\textbf{$\widehat{X}$}}

\begin{document}
\begin{titlepage}
\title{Faster generation of random spanning trees}

\thispagestyle{empty}
\author{Jonathan A. Kelner\thanks{Research partially supported by NSF grant CCF-0843915.} \\ Massachusetts Institute of Technology \\ kelner@mit.edu 
\and
Aleksander M\k{a}dry\thanks{Research partially supported by  Fulbright Science and Technology Award, by NSF contract CCF-0829878 and by ONR grant N00014-05-1-0148.} \\Massachusetts Institute of Technology \\madry@mit.edu}

\date{}

\maketitle
\input abstract

\thispagestyle{empty}
\end{titlepage}

\input preliminaries

\input decomposition

\input gentree

\input decomposition_alg

\input improving

\bibliographystyle{abbrv}
\bibliography{references}

\end{document}

%% file: abstract.tex
\begin{abstract}
In this paper, we set forth a new algorithm for generating approximately uniformly random spanning trees in undirected graphs.  We show how to sample from a distribution that is within a multiplicative $(1+\delta)$ of uniform in expected time $\TO(m\sqrt{n}\log 1/\delta)$.  This improves the sparse graph case of the best previously known worst-case bound of $O(\min \{mn, n^{2.376}\})$, which has stood for twenty years.

To achieve this goal, we exploit the connection between random walks on graphs and electrical networks, and we use this to
 introduce a new approach to the problem that integrates discrete random walk-based techniques with continuous linear algebraic methods.  We believe that our use of electrical networks and sparse linear system solvers in conjunction with random walks and combinatorial partitioning techniques is a useful paradigm that will find further applications in algorithmic graph theory.
 \end{abstract}

%% file: preliminaries.tex
\section{Introduction}
In this paper, we set forth a new algorithm for generating random spanning trees in undirected graphs.
Random spanning trees are among the oldest and most extensively investigated probabilistic objects in graph theory, with their study dating back to Kirchoff's work in the 1840s \cite{Kirchoff}.
However, it is only in the past several decades that researchers have taken up the question of how best to generate uniformly random spanning trees algorithmically.
This question became an active area of research in the 1980s and 1990s, during which a long string of papers appeared that provided successively faster algorithms for this task (e.g.,\cite{Gu'enoche83randomspanning,Kulkarni90generatingrandom, Colbourn88estimatingthe, Colbourn, Broder, Aldous90, Kandel96shufflingbiological, Wilson96generatingrandom}).

Previous algorithms for this problem broadly fall into two categories: determinant-based algorithms and random walk-based algorithms.  The starting point for the determinant-based algorithms was Kirchoff's Matrix Tree Theorem, which reduces counting the number of spanning trees in a graph to the evaluation of a determinant \cite{Kirchoff} (see, e.g., Ch.2,  Thm. 8 in \cite{Bollobas79graphtheory:}).  The first such algorithm produced a random spanning tree in time $O(mn^3)$ \cite{Gu'enoche83randomspanning, Kulkarni90generatingrandom}.  After sequence of improvements (\cite{Colbourn88estimatingthe, Colbourn89unrankingand}), this line of research culminated in the algorithm of Colbourn, Myrvold, Day, and Nel \cite{Colbourn}, which runs in the amount of time necessary to multiply two $n\times n$ matrices, the best known bound for which is $O(n^{2.376})$ \cite{Coppersmith90matrixmultiplication}.

The random walk-based algorithms began with the following striking theorem due to Broder \cite{Broder} and Aldous \cite{Aldous90}:

\begin{theorem}\label{thm:randwalkalg}
Suppose you simulate a random walk in an undirected graph $G=(V,E)$, starting from an arbitrary vertex $s$ and continuing until every vertex has been visited.  For each vertex  $v\in V\setminus\{s\}$, let $e_v$  be the
  edge through which $v$ was visited for the first time in this walk.  Then,  $T=\{e_v\ | v\in V\setminus\{s\}\}$ is a uniformly random 
spanning tree of $G$.
\end{theorem}

This immediately yields an algorithm for generating a random spanning tree whose running time is proportional to the cover time of $G$.  If $G$ has $n$ vertices and $m$ edges, the cover time can be $\Theta(mn)$ in the worst case, but it is often much smaller.  For sufficiently sparse graphs, this yields a better worst-case running time than the determinant-based algorithms.
Since one clearly needs to see every vertex in $G$, it would seem unlikely that such methods could run in less than the cover time of the graph.  However, in the last major breakthrough in this line of research, Wilson \cite{Wilson96generatingrandom} showed that, by using a different random process, one could actually generate spanning trees in expected time proportional to the mean hitting time of the graph, which can be much smaller than the cover time (but has the same worst-case asymptotics).  These algorithms generate an exactly uniform random spanning tree, but they remain the best known algorithms even if one wants to generate a spanning tree from a distribution that is within some multiplicative $(1+\delta)$ of uniform.  (We will call this a $\delta$-random spanning tree; we shall define it more precisely in section \ref{sec:cond_rand}.)

The worst-case running time bound of $O(mn)$ has stood for twenty years.  In this paper, our main result is a new algorithm that offers a better worst-case time bound:
\begin{theorem}\label{thm:main}
Let $G$ be a graph with $n$ vertices and $m$ edges. For any $\delta>0$, we can generate a $\delta$-random spanning tree of $G$ in expected time
$\TO(m\sqrt{n}\log 1/\delta)$.
\end{theorem}


Beyond the classical applications of generating random spanning trees that motivated the original work on the problem, there have been some developments that further motivate their study.  In particular, a recent paper of  Goyal, Rademacher, and Vempala~\cite{GoyalRV09} showed how to use random spanning trees to generate efficient sparsifiers of a graph, and they then explained how this could be used to provide a scalable and robust routing scheme.

We believe that our techniques are of independent interest and may provide a good set of tools for the solution of other problems. On a broad level, we would like to highlight the use of electrical flows and linear systems to approach combinatorial questions about graphs.  In addition to our work, they were recently used by Spielman and Srivastava~\cite{SpielmanS_electricalresistances}, and Batson, Spielman and Srivastava \cite{SpielmanB} to provide much stronger sparsifiers than were previously available.
These and the the present work illustrate two distinct ways in which electrical flows on a graph provide much richer information than that conveyed by just the graph spectrum.  The fact that they may be found in nearly-linear time~\cite{SpielmanTeng_solving} provides a powerful tool for answering questions about graphs.  To our knowledge, this is the first use of such techniques in combination with combinatorial ones to provide a faster algorithm for a purely combinatorial question.  We believe that this is an exciting new frontier of algorithmic spectral graph theory that will find many further applications.

In addition, it is interesting to note that our algorithm takes a graph problem for which fast matrix multiplication provides the best known methods for dense graphs and uses a method based on a sparse linear system solver to obtain a better running time for sparse graphs.  There are a large number of graph problems for which fast matrix multiplication provides the best known running time.  Our techniques suggest a general template for using the linear algebraic intuition underlying these algorithms in conjunction with sparse linear algebra routines to obtain faster algorithms on sparse graphs.

\subsection{$\delta$-random spanning trees and arborescences}\label{sec:cond_rand}

Formally, we consider the following problem: given an undirected graph $G=(V,E)$ with $n$ vertices and $m$ edges, find a randomized algorithm $A$ that, for each spanning tree $T$ of $G$,  outputs $T$ with probability $1/|\mathcal{T}(G)|$, where $\mathcal{T}(G)$ is the set of all spanning trees of $G$. We will be interested in a slightly relaxed version of this problem in which we require $A$ to output each tree $T$ with a probability $p(T)$ that is {\em $\delta$-far} from the uniform i.e.  $(1-\delta)/|\mathcal{T}(G)|\leq p(T) \leq (1+\delta)/|\mathcal{T}(G)|$, for some parameter $\delta>0$.  We call this {\it generating a $\delta$-random spanning tree}.  We note that our algorithms' dependence on $\delta$ occurs only because some of our algebraic manipulations are carried out only to finite precision.  As such, we depend only \emph{logarithmically} on $1/\delta$, not polynomially.

For technical reasons that will arise later in the paper, it will be useful to consider arborescences in addition to trees.  For a given $s\in G$, an {\em arborescence $T$ rooted at $s$} is a directed spanning tree of $G$ in which all vertices in $G\setminus \{s\}$ have exactly one incoming arc. We use the notation $r(T)$ to denote the root of an arborescence, and we use  $e_T(v)$, for any $v\in G\setminus \{r(T)\}$, to denote the unique arc incoming to $v$ in $T$.

We say that a procedure $A$ {\em generates a conditionally random arborescence} if it outputs a vertex $s$ and an arborescence $T$ rooted at $s$ such that the probability, conditioned on some $s$ being a root, of outputting a particular arborescence $T$ rooted at $s$ is $1/|\mathcal{T}_s(G)|$, where $\mathcal{T}_s(G)$ is the set of all arborescences in $G$ rooted at $s$.\footnote{For brevity, we will hereafter omit the word ``conditionally'' when we refer to such an object.  We stress that this is an object that we are introducing for technical reasons, and it should not be confused with the different problem of generating a uniformly random arborescence on a \emph{directed} graph.}  Note that in this definition we do not make any restrictions on  the probabilities $p_A(s)$ that the generated arborescence output by $A$  is rooted at $s$.  Now, it is easy to see that, once we  fix some $s\in G$, there is one-to-one correspondence between spanning trees of $G$ and arborescences rooted at $s$. Indeed, given any spanning tree, there is a unique way of directing its edges to make it a valid arborescence rooted at $s$; conversely, given any arborescence rooted at $s$, we can obtain a spanning tree by just disregarding the direction of the edges. As a result, if we have a procedure $A$ that generates a random arborescence  then, for a given spanning tree $T$, the probability that it will be generated is exactly $\sum_{s\in G} p_A(s)/|\mathcal{T}_s(G)|=\sum_{s\in G} p_A(s)/|\mathcal{T}(G)|=1/|\mathcal{T}(G)|$. This means that if we interpret the arborescence returned by $A$ as a spanning tree then we get in this way a random spanning tree. By completely analogous reasoning, we get that a procedure $A$ that generates $\delta$-random arborescences gives us a procedure that generates $\delta$-random spanning trees.

\subsection{An outline of our approach}

To describe our approach, let us consider a random walk $X$ in $G$ whose starting vertex is chosen according to the stationary distribution of $G$. If we just simulate $X$ step-by-step until it covers the whole graph $G$, Theorem \ref{thm:randwalkalg} asserts that from a transcript $X(\omega)$ of this simulation we can recover a random arborescence that is rooted at the starting vertex $s(\omega)$ of $X(\omega)$. However, while the whole transcript can have an expected length of  $\Omega(mn)$, we utilize only a tiny fraction of entries---the $O(n)$ entries that allow us to recover arcs $e_v$ for $v\in V\setminus \{s(\omega)\}$.  It is thus natural to wonder whether the generation of the whole transcript is necessary.
The random walk may spend long periods of time walking around regions of the graph that have already been covered, which seems quite wasteful.  One may ask whether it is possible to identify and avoid such situations by somehow skipping such unnecessary steps.
That is, one may ask whether there is a way of shortcutting the walk $X$ such that the corresponding transcripts are much shorter, can be generated efficiently,  and still retain the information that we need to recover the desired arborescence.   We note that this intuition is quite reasonable for many of the standard examples of graphs
that have large cover time, which consist of regions that are covered very quickly but in which the walk spends much of its time.

A tempting way to obtain such a shortcutting would be to try to just cut out from $X$ all of its parts that correspond to visiting already explored parts of $G$. This shortcutting yields transcripts of length $O(n)$ and contains all of the information that we need. Unfortunately, it is not clear whether an efficient way of generating such transcripts exists---it is quite possible that the fastest way to find the next edge to an unvisited vertex traversed by the walk is to generate the whole trajectory of $X$ inside the previously visited parts of $G$  step-by-step.

The core of our approach is showing that there indeed exists an efficient way of shortcutting $X$. On a high level, the way we obtain it is as follows. We start by identifying a number of induced subgraphs $D_1,\ldots ,D_k$ of $G$ such that the cover time of each $D_i$ is relatively small, and the set $C$ of edges of $G$ that are not inside any of the $D_i$s constitutes a small fraction of the edges of $G$. Now, we shortcut the walk $X$ in $G$ by removing, for each $i$, the trajectories of $X$ inside $D_i$ that occur after $X$ has already explored {\em the whole} $D_i$.\footnote{Later we modify this procedure slightly to get better running time for dense graphs.} Such shortcutted transcripts clearly retain all of the information that we need. Moreover, we show that the fact that $D_i$s have small cover time and $C$ has very small size imply that the expected length is also small. Finally, by exploiting the connection between random walks and electrical flows together with the linear system solver of Spielman and Teng \cite{SpielmanTeng_solving},  we provide an efficient procedure to approximately generate such shortcutted transcripts of $X$. All together, this yields the desired procedure for generating $\delta$-random arborescences.

\subsection{Outline of the paper}
The outline of the paper is as follows. In section \ref{sec:structure}, we formally define the decompositions of $G$ in which we are interested. We then relate the structure of the random walk $X$ to these decompositions. Next, in section \ref{sec:our_algorithm}, we show how these ideas can be used to develop an algorithm that generates (conditionally) $\delta$-random arborescence in expected time $\TO(m^2/\sqrt{n}\log 1/\delta)$. Finally, in section \ref{sec:improving}, we prove Theorem \ref{thm:main} by refining the previous algorithm to make it run in expected time $\TO(m\sqrt{n}\log 1/\delta)$.

%% file: decomposition.tex
\section{ The structure of the walk $X$}\label{sec:structure}

In this section, we shall formally define the decomposition of $G$ that will be the basis of our algorithm and prove several important facts about the structure of the walk $X$ with respect  to this decomposition.

\subsection{$(\phi,\gamma)$-decompositions}

While the algorithm sketched in the previous section could be made to use only edge cuts, our faster algorithm in section~\ref{sec:improving} will require both vertex and edge cuts.  To facilitate this, we shall use a decomposition that permits us to keep track of a both a set of edges $C$ and a set of vertices $S$ in the cut.

To this end, let  $(D_1,\ldots, D_k,S,C)$ denote a partition of $G$ such that $S\subseteq V(G)$, $\bigcup_i V(D_i) = V(G)\setminus S$, the $D_i$ are disjoint induced subgraphs of $G$, and $C=E(G)\setminus \bigcup_i E(D_i)$ is the set of edges not entirely contained inside one of the $D_i$.
 For a given $D_i$, let $U(D_i)$ be the set of vertices of $D_i$ incident to an edge from $C$, and let $C(D_i)$ be the subset of $C$ incident to $D_i$.

\begin{definition}\label{def:decomp1}
A partition $(D_1,\ldots, D_k,S,C)$ of $G$ is a {\em ($\phi$, $\gamma$)-decomposition} of $G$ if:
\begin{enumerate}
\item $|C|\leq \phi |E(G)|$,
\item  for each $i$, the diameter $\gamma(D_i)$ of $D_i$ is less than or equal to $\gamma$, and
\item  for each $i$, $|C(D_i)|\leq |E(D_i)|$.
\end{enumerate}
\end{definition}

 Note that this definition does not constrain the size of $S$ explicitly, but $|S|$ is implicitly bounded by the fact that all edges incident to vertices of $S$ are included in $C$.

 Intuitively, the above decomposition corresponds to identifying in $G$ induced subgraphs $D_1,\ldots ,D_k$ that have diameter at most $\gamma$ and contain all but a $\phi$ fraction of edges of $G$.
 We bound the diameters of the $D_i$s and ensure (by the third condition) that they are not too  ``spread out''  in $G$, since these properties will allow us to prove that $X$ covers each of them relatively quickly. We will do this using the approach of  Aleliunas {\em et al.} \cite{Aleliunas} that proved that the cover time of an unweighted graph $G'$ with diameter $\gamma(G')$ is at most $O(|E(G')|\gamma (G'))$.

 For now,  let us assume that we have been given some fixed $(\phi, \gamma)$-decomposition $(D_1,\ldots, D_k,S,C)$ of $G$ (for some $\gamma$ and $\phi$ that we will determine later).
 In Lemma~\ref{lem:decomp}, we will show that such decompositions with good parameters exist and can  be constructed efficiently.

\subsection{ The walk $X$}


Let $X=(X_i)$ be a random walk in $G$ that is started at a vertex chosen according to the stationary distribution of $G$, where $X_i$ is the vertex visited in the $i$-th step.  
Let $\tau$ be the
 time corresponding to the first moment when our walk $X$ has visited all of the vertices of $G$. Clearly, $E(\tau)$ is just the expected cover time of $G$, and by the fact that $G$ has $m$ edges and diameter at most $n$,  the result of Aleliunas et al. \cite{Aleliunas} yields
\begin{fact}\label{fa:length_of_covering_walk}
$E[\tau]=O(mn).$
\end{fact}


Let $Z$ be the random variable corresponding to the number of times that some edge from $C$ was traversed by our random walk $X$, i.e., $Z$ is the number of $i< \tau$ such that $X_i=v$, $X_{i+1}=v'$, and $(v,v')\in C$. Since, by Fact \ref{fa:length_of_covering_walk}, the expected length of $X$ is $O(mn)$, and we choose the starting vertex of $X$ according to stationary distribution of $G$, the expected number of traversals of edges from $C$ by $X$ is just proportional to its size, so, on average, every $m/|C|$-th step of $X$ corresponds to an edge from $C$. Therefore,  the fact that in our decomposition $|C|\leq \phi |E(G)|$ implies:

\begin{fact}\label{fa:length_of_c}
$E[Z]=O(\phi mn)$.
\end{fact}

Let $Z_i$ be the random variable corresponding to the number of times that some edge inside $D_i$ is traversed by our walk. By definition, we have that $\tau=\sum_i Z_i+Z$. Now, for each $D_i$, let $\tau_i$ be the time corresponding to the first moment when we reach some vertex in $U(D_i)$ {\em after} all the vertices from $D_i$ have been visited by our walk $X$.

Finally, let $Z_i^*$ be the random variable corresponding to the number of times that some edge from $E(D_i)$ is traversed by our walk $X$ until
time $\tau_i$ occurs, i.e., until $X$ explores the whole subgraph $D_i$. The following lemma holds:

\begin{lemma}\label{lem:induced_graph}
For any $i$, $E[Z_i^*]=\TO(|E(D_i)|\gamma(D_i))$.
\end{lemma}

Before we proceed to the proof, it is worth noting that the above lemma does not directly follow from result of Aleliunas {\em et al.}\cite{Aleliunas}. The reason is that \cite{Aleliunas} applies only to a natural random walk in $D_i$, and the walk induced by $X$ in $D_i$ is different. Fortunately, the fact that
$|C(D_i)|\leq |E(D_i)|$ allows us to adjust the techniques of \cite{Aleliunas} to our situation and prove that a bound similar to the one of Aleliunas {\em et al.} still holds.

\begin{Proof}
Let us fix $D=D_i$.
For a vertex $v\in V(D)$, let $d_G(v)$ be the degree $v$ in $G$, and let $d_D(v)$ be the degree of $v$ in $D$.
Clearly, $d_G(v)\geq d_D(v)$, and $d_C(v)\equiv d_G(v)-d_D(v)$ is the number of edges from $C$ incident to $v$.
For $u,v \in U(D)$, let $p^D_{u,v}$ be the probability that  a random walk in $G$ that starts at $u$ will reach $v$ through a path that does not pass through any edge inside $D$.

Consider a (weighted) graph $D'$, which we obtain from $D$ by adding, for each $u,v\in U(D)$, an edge $(u,v)$ with weight $d_G(u)\cdot p^D_{u,v}$.  (All edges from $D$ have weight $1$ in $D'$.  Note that we do not exclude the case $u=v$, so we may add self-loops.)
By
 the fact that the Markov chain corresponding to the random walk $X$ in $G$ is reversible, $d_G(u)\cdot p^D_{u,v}=d_G(v)\cdot p^D_{v,u}$, so our weights are consistent.

Now, the crucial thing to notice is that if we take our walk $X$ and filter out of the vertices that are not from $D$, then the resulting ``filtered'' walk $Y_D$ will just be a natural random walk in $D'$ (with an appropriate choice of the starting vertex). In other words, the Markov chain corresponding to the random walk in $D'$ is exactly the Markov chain induced on $V(D)$ by the Markov chain described by our walk $X$. As a result, since $Z_i^*$ depends solely on the information that the induced random walk $Y_D$ retains from $X$, it is sufficient to bound $Z_i^*$ with respect to $Y_D$. However, it is easy to see that, in this case, $E[Z_i^*]$ can be upper-bounded by the expected time needed by a random walk in $D'$, started at arbitrary vertex, to visit all of the vertices in $D'$ and then reach some vertex in $U(D)$. We thus can conclude that $E[Z_i^*]$ is at most twice the cover time of $D'$.  (More precisely, it is the cover time plus the maximum hitting time.)

Now, it is well-known (e.g., see \cite{Lov_randomwalks}) that the cover time of any undirected graph $G'$ is at most $2\log |V(G')| H_{max}$, where $H_{max}(G')$ is the maximal hitting time in $G'$. Our aim is therefore to show that $H_{max}(D')=O(|E(D)|\gamma(D))$,
where $\gamma(D)$ is the diameter of $D$. We will do this by applying the approach of Aleliunas {\em et al.}, who proved in \cite{Aleliunas} that for an {\em unweighted} graph $G'$, $H_{max}(G')\leq |E(G')| \gamma(G')$.

To achieve this goal, let $K$ be some number such that all weights in $D'$ after multiplication by $K$ become integral (which we can do since all of the weights in $D'$ have to be rational). Let $K\cdot D'$ be the {\em unweighted}  multigraph that we obtain from $D'$ by multiplying all of its weights by $K$ and then interpreting the new weights as multiplicites of edges. Note that the natural random walk in $K\cdot D'$ (treated as a sequence of visited vertices) is exactly the same as in $D'$.

Now, we prove that for two vertices $v$ and $w$ of $D'$ such that an edge $(v,w)$ exists in $D$,
the expected time $H_{v,w}(K\cdot D) = H_{v,w}(D')$
until a random walk in $K\cdot D'$ that starts at $v$ will reach $w$ is at most $4|E(D)|$. To see this, note that the long-run frequency with which an copy of an edge is taken in a particular direction is $1/(2M)$, where $M$ is total number of edges of $K\cdot D'$ (and we count each copy separately). Thus one of the $K$ copies of edge $(v,w)$ is taken in the direction from $v$ to $w$ every $K/(2M)$-th step on average. This in turn means that $H_{v,w}(D')\leq 2M/K$. Now, to bound $M$, we note first that $M\leq K (|E(D)|+\sum_{u,v\in U(D)} d_G(u) p^D_{u,v})$. Thus, since for a given $u\in U(D)$ the probability $\sum_{v} p^D_{u,v}$ of going outside $D$ directly from $u$ is equal to $d_C(u)/d_G(u)$, we obtain that $M\leq K(|E(D)|+ \sum_u d_C(u))\leq 2 K |E(D)|$ by the property of a $(\phi,\gamma)$-decomposition that requires that $|C(D)|\leq |E(D)|$. We can thus conclude that $H_{v,w}(D')\leq 2M/K \leq 4 |E(D)|$, as desired. Having obtained this result, we can use a simple induction to show that $H_{v,w}(D')\leq 4|E(D)|\Delta(v,w)$, where $\Delta(v,w)$ is the distance between $v$ and $w$ in $D$. From this, we can conclude that $H_{max}(D')=4|E(D)|\gamma(D)$ and $E[Z_i^*]\leq 16 \log n |E(D_i)|\gamma(D_i)$, as desired.

\end{Proof}

%% file: gentree.tex
 
\section{Our algorithm}\label{sec:our_algorithm}

Let us now focus our attention on some particular $D_i$ from our $(\phi, \gamma)$-decomposition of $G$. The idea of our algorithm is based upon the following observation. Suppose we look at our random walk $X$ just after time $\tau_i$ occurred. Note that this means that we already know for all $v\in V(D_i)$ which arc $e_v$ we should add to the final arborescence. Therefore, from the point of view of building our arborescence, we gain no more information by knowing
what trajectory $X$ takes inside $D_i$ after
time $\tau_i$. More precisely, if, at some step $j$, $X$ enters $D_i$ through some vertex $v\in V(D_i)$ and, after $k$ steps, leaves through some edge $(u,u')\in C$, where $u\in V(D_i)$ and $u'\notin V(D_i)$,
the actual trajectory $X_j,\ldots,X_{j+k}$ does not matter to us.
The only point of simulating $X$ inside $D_i$ after time $\tau_i$ is to learn, upon entering $D_i$ through $v$, through which edge $(u,u')\in C$ we should leave.

Let $P_v(e)$ be the probability of $X$ leaving $D_i$ through $e$ after entering through vertex $v$.  If we knew $P_v(e)$ for all $v\in V(D_i)$ and all $e\in C(D_i)$,
then we could just, upon entering $v$, immediately choose the edge $e$ through which we will exit according to distribution $P_v(e)$ without computing the explicit trajectory of $X$ in $D_i$. That is, if we consider a shortcutting $\TX$  of $X$ that cuts out from $X$  all trajectories inside $D_i$ after it was explored, then $P_v(e)$ would be all that we need to simulate $\TX$ in time proportional to the length of $\TX$ (as opposed to the length of $X$).

Now, the point is that we can efficiently compute an $\varepsilon$-approximation of $P_v(e)$, as we will show in section~\ref{subsec:computing_p}, which will enable us to compute this shortcutting to fairly high precision.
Furthermore, as we will see shortly, the analysis of structure of $X$ that we performed in the previous section shows that the computation of these 'unnecessary' trajectories constitutes the bulk of the work involved in a faithful simulation of $X$, and therefore $\TX$ has much smaller length while yielding the same distribution on random arborescences.

To formalize the above intuitions, let $\TX=(\TX_i)$ be a random walk obtained in the following way. Let $X(\omega)=X_0(\omega),\ldots X_{\tau(\omega)}(\omega)$ be some concrete trajectory of $X$, and let $X(\omega)=\overline{X}_1(\omega), \ldots ,\overline{X}_{k(\omega)}(\omega)$ be decomposition of $X(\omega)$ into contiguous blocks $\overline{X}_j(\omega)$ that are contained inside $D_{i_j}$ for some sequence $i_j\in \{0,\ldots ,k\}$, where we adopt the convention that $D_0=S$. We define $\TX(\omega)$ as follows. We process $X(\omega)$ block-by-block and we copy a block $\overline{X}_j(\omega)$ to $\TX(\omega)$ as long as $\tau_{i_j}$ has not occurred yet or $i_j=0$, otherwise we copy to $\TX(\omega)$ only the first and last entries of the block.  (We shall refer to the latter event as a {\em shortcutting of the block}.) We proceed now to the formal proofs of the properties of $\TX$ stated above.

We start by showing that $\TX$ can indeed be simulated efficiently given probabilities $P_v(e)$ and some preprocessing time that will later be amortized into the overall running time of the algorithm:

\begin{lemma}\label{lem:datastructure}
Knowing $P_v(e)$ for all $e\in C(D_i)$, $v\in V(D_i)$ and $i$, we can preprocess these values in $\TO(\phi mn)$ time in such a way that it allows simulation of $l$ steps of $\TX$ in time $\TO(l)$.
\end{lemma}

\begin{Proof}
Simulating $\TX$ before any $\tau_i$ has occurred is straightforward. The only  thing to show is that we can implement the shortcutting of the blocks, for which it would suffice to show that, using the $P_v(e)$, we can sample an edge $e$ from this distribution in polylogarithmic time. To see how this can be done, let us fix some $i$ and some ordering $e_1,\ldots ,e_{|C(D_i)|}$ of the edges from $C(D_i)$. For each $v\in U(D_i)$, we can create in $\TO(|C(D_i)|)$ time an array $A_v(i)$ where $A_v(i)=\sum_{1\leq j\leq i} P_v(e_j)$,
so we can do it for all $v\in U(D_i)$ in total time at most $\TO(|V(D_i)|\cdot |C(D_i)|)$.

Now, if we want to choose $e$ according to $P_v(e)$, we choose a random number from $r\in[0,1]$, use binary search to find an entry $j$ in $A_v(i)$ such that $A_v(j)\geq r>A_v(j-1)$, and output $e=e_j$. Summing up the total preprocessing time for all $D_i$ yields the desired bound.
\end{Proof}

We are ready to prove the following lemma:

\begin{lemma}\label{lem:wrapup1}
Given an $(\phi,\gamma)$-decomposition of $G$ and the probabilities $P_v(e)$, we can find a random arborescence of $G$ in expected time $\TO(m(\gamma+\phi n))$.
\end{lemma}

\begin{Proof}
By the discussion above, it suffices to simulate the algorithm described in Theorem~\ref{thm:randwalkalg} using the walk $\TX$, so we just need to bound the expected number of steps it takes for $\TX$ to cover the whole graph $G$.
To do this, we upper-bound the expected length of the covering walk $X$ after shortcutting it.  However, this quantity is just $\sum_i E[Z_i^*]+3E[Z]$, since the two vertices that remain in $\TX$ after shortcutting some block from $X$ can be amortized into the number of traversals by $X$ of edges from $C$. By Fact \ref{fa:length_of_c} and Lemma \ref{lem:induced_graph}, we get that $\sum_i E[Z_i^*]+3E[Z]=\TO(\sum_i |E(D_i)|\gamma + \phi mn)=\TO(m(\gamma+\phi n))$.
By Lemma~\ref{lem:datastructure}, we can simulate a walk of this length in expected time $\TO(m(\gamma+\phi n))$.
\end{Proof}

In order to complete our algorithm, we thus need to show three things: we can quickly compute the $P_v(e)$, a $(\phi,\gamma)$-decomposition exists with good parameters, and we can find such a decomposition efficiently.

\subsection{Computing $P_v(e)$}
\label{subsec:computing_p}

In the following lemma we address the first issue.

\begin{lemma}\label{lem:probabilities}
Given a $(\phi,\gamma)$-decomposition of $G$, we can compute multiplicative $(1+\varepsilon)$-approximations of all of the $P_v(e)$ in time $\TO(\phi m^2\log 1/\varepsilon)$.
\end{lemma}

\begin{Proof}
Let us fix some $D=D_i$ and an edge $e=(u,u')\in C(D)$ with $u\in U(D)$. Consider now a graph $D'$
that we obtain from $D$ as follows. First, we add vertex $u'$, and some dummy vertex $u^*$ to $D$, and then, for
each $(w,w')\in C(D)\setminus\{e\}$ with $w\in U(D)$, we add an edge $(w,u^*)$.  (Note that $w'$ can be
equal to $u'$.) Finally, we add the edge $e=(u,u')$. The crucial thing to notice now is that for any
given vertex $v\in D$, $P_v(e)$ is exactly the probability that a random walk in $D'$ started at
$v$ will hit $u'$ before it hits $u^*$.  We can compute such probabilities quickly using
electrical flows.

More precisely, (see, e.g., \cite{Lov_randomwalks}) if we treat $D'$ as an electrical circuit in which
we impose voltage of $1$ at $u'$ and $0$ at $u^*$, then the voltage achieved at $v$ in such an
electrical flow is equal to $P_v(e)$. We can compute a $(1+\varepsilon)$-approximation of such a
flow in time $\TO(|E(D')|\log 1/\varepsilon)$  using the
linear system solver of Spielman and Teng~\cite{SpielmanTeng_solving}.
To do so, let $L$ be the Laplacian of $D'$, where we let the first two rows correspond to $u'$ and $u^*$, respectively.
Furthermore, let  $\mathbf{i}_{ext}\in \mathbb{R}^{|V(D')|}$ be the vector that is $1$ in its first coordinate, $-1$ in its second coordinate, and zero everywhere else.
Let $\mathbf{v'}\in \mathbb{R}^{|V(D')|}$ be the solution to the linear system $L \mathbf{v}' =\mathbf{i}_{ext}$, which we can approximately find in nearly-linear time using the solver from~\cite{SpielmanTeng_solving}.  We obtain our desired vector of voltages $\mathbf{v^*}$ from $\mathbf{v'}$ by subtracting $v_{2}'$ from all of its coordinates and dividing them by $v_{1}'-v_{2}'$. It is worth noting that this approach to computing electrical flow was also used in Spielman and Srivastava \cite{SpielmanS_electricalresistances}.

Our algorithm computes such voltages for each edge $e$.
(Note that we might have to compute two such voltages per edge---one for each endpoint of $e$---if they each  lie in different $D_i$.)
From each flow, we store the probabilities $P_v(e)$ for all vertices $v$ that we are interested in.
The running time of such procedure is bounded by $\TO(|C|\sum_i |E(D_i)|\log
1/\varepsilon)=\TO(\phi m^2\log 1/\varepsilon)$, where we use the fact that for each $D$, $|E(D')|=|E(D)|+|C(D)|\leq 2|E(D)|$ by the definition of a $(\phi,\gamma)$-decomposition.
\end{Proof}

Since the linear system solver only gives us approximate values for the $P_v(e)$, we need to show that we can control the overall error while still maintaining a good running time.  We do this with the following lemma:

\begin{lemma}\label{conj}
Given a $(\phi,\gamma)$-decomposition of $G$ and multiplicative $(1+\varepsilon)$-approximations of all of the probabilities $P_v(e)$, we can generate a $\delta$-random
arborescence of $G$ in expected time $\TO(m(\gamma+\phi n))$, as long as $\varepsilon\leq \delta /mn$.
\end{lemma}

\begin{Proof}

We define a new random walk  $\TX'$ that approximates $\TX$ as follows.  The walk $\TX'$ simulates $\TX$ for the first $mn$ steps, except it uses the approximate values of the $P_v(e)$ when it shortcuts a block. After the $mn^\text{th}$ step, it shortcuts blocks using the exact values of the $P_v(e)$.


We start by noting that, for any $t\geq 0$ and trajectory $\TX(\omega)$,
\begin{align}
\Pr[&\TX_0=\TX(\omega)_0,\ldots ,\TX_t=\TX(\omega)_t]/(1+\delta) \nonumber\\ 
&\leq  \Pr[\TX_0'=\TX(\omega)_0,\ldots ,\TX_t'=\TX(\omega)_t]  \label{eq:delta_prob}\\ 
&\leq  (1+\delta)\Pr[\TX_0=\TX(\omega)_0,\ldots ,\TX_t=\TX(\omega)_t].  \nonumber
\end{align}

That is, for any trajectory $\TX(\omega)$, the probability that its first $t$ entries appear as a prefix of the walk $\TX'$ is within a $(1+\delta)$ multiplicative factor of the probability that this prefix appears as a prefix of the walk $\TX$.

Indeed, each of the first $mn$ steps of the walk $\TX'$ distorts the likelihood of appearance of a particular sequence of vertices by a multiplicative factor of $(1+\varepsilon)$. Thus, since $(1+\varepsilon)^{mn}\leq 1+\varepsilon mn\leq 1+\delta$, the assertion follows. 

It is easy to see that (\ref{eq:delta_prob}), together with  Theorem \ref{thm:randwalkalg} and the fact that the walk $\TX$ is a shortcutting of the walk $X$, implies that simulating the walk $\TX'$ until it covers the graph $G$ is sufficient to obtain a $\delta$-random arborescence. Therefore, what is left to prove is that the simulation of the walk $\TX'$ until it covers the graph $G$ can be done in expected time $\TO(m(\gamma+\phi n))$. 

We do this as follows. For the simulation of the first $mn$ steps, we use the method from Lemma \ref{lem:datastructure} with the $(1+\varepsilon)$-approximations of the probabilities $P_v(e)$. If the walk hasn't covered the graph by step $mn$, we continue the simulation of $\TX'$ by simply generating  the entire transcript of the walk $X$ until it covers the graph and then shortcutting it accordingly. 

To bound the running time of this procedure, we note that  (\ref{eq:delta_prob}) implies
that the expected number of steps necessary for $\TX'$ to cover $G$ is at most $(1+\delta)$ times the expected number of steps required by $\TX$,
%
%
 which, by Fact \ref{fa:length_of_c} and Lemma \ref{lem:induced_graph}, is $\sum_i E[Z_i^*]+3E[Z]=\TO(m(\gamma+\phi n))$. Therefore, by Lemma \ref{lem:datastructure}, we get that the part of the simulation that deals with the first $mn$ steps runs in expected time $\TO((1+\delta) \sum_i E[Z_i^*]+3E[Z])=\TO(m(\gamma+\phi n))$, since we can always assume that $\delta\leq 1$. 
 
 By the Markov inequality, we know that the probability that $\TX'$ ever takes more than $mn$ steps is at most $\TO(m(\gamma+\phi n))/mn$. We can bound the expected running time of the second part of the simulation by this probability multiplied by the the expected cover time of the whole graph, which is $O(mn)$. This gives the total expected running time of the simulation to be $\TO( m(\gamma+\phi n)+m^2n(\gamma+\phi n)/mn)=\TO(m(\gamma+\phi n))$, as desired.

\end{Proof}

We now proceed to the final ingredient of our algorithm---finding good $(\phi,\gamma)$-decompositions quickly.

%% file: decomposition_alg.tex
\subsection{Obtaining good ($\phi$, $\gamma$)-decompositions quickly}

By using the ball-growing technique of Leighton and Rao \cite{LeightonRao},  we obtain the following:

\begin{lemma}\label{lem:decomp}
For any $G$ and any $\phi=o(1)$, there exists a $(\phi, \TO(1/\phi))$-decomposition of
$G$. Moreover, such a decomposition can be computed in time $\TO(m)$.
\end{lemma}
\noindent We omit the proof, as we shall prove a stronger statement in Lemma~\ref{lem:decomp2}.

\subsection{Generating $\delta$-random arborescence in expected time $\TO(m^2/\sqrt{n}\log 1/\delta)$}

We can now put these results together
to generate $\delta$-random arborescences in expected time $\TO(m^2/\sqrt{n}\log 1/\delta)$.
Note that this bound is worse than the one stated in Theorem \ref{thm:main}---we shall improve it to obtain the better time bound in section \ref{sec:improving}.

 Let   $\phi=1/n^{1/2}$, and let
$\varepsilon=\delta/mn$, as in Lemma \ref{conj}. By Lemma \ref{lem:decomp}, we can get a
$(1/n^{1/2},\TO(n^{1/2}))$-decomposition of $G$ in $\TO(m)$ time. By Lemma
\ref{lem:probabilities}, we can compute the estimates of $P_v(e)$ for all relevant $v$ and $e$
in time $\TO(m^2/\sqrt{n}\log 1/\varepsilon)=\TO(m^2/\sqrt{n}\log 1/\delta)$.
Having done this, we can use
Lemma \ref{lem:wrapup1} (together with Lemma \ref{conj}) to generate a
$\delta$-random arborescence in time $\TO(m^2/\sqrt{n}\log 1/\delta)$.

%% file: improving.tex
\section{Obtaining an $\TO(m\sqrt{n}\log 1/\delta)$ running time}
\label{sec:improving}
The bottleneck in the procedure presented above is the computation of probabilities $P_v(e)$---everything else can be done in time $\TO(m\sqrt{n}\log
1/\delta)$. Unfortunately, it is not clear how we could improve the running time of these computations. To circumvent this problem, we will alter our procedure to use slightly different probabilities and a slightly different random walk that will end up yielding a faster simulation time.

To introduce these probabilities, let us assume that there are no edges in $G$ between different $D_i$s (which we will ensure to be the case), and let $C_i(u)$ for a vertex $u\in S$ be the set of edges incident both to $u$ and the component $D_i$. Now, for a given $i$, some $v\in V(D_i)$, and $u\in S$ with $|C_i(u)|>0$, we define $Q_v(u)$ to be the probability that $u$ is the first vertex not in $V(D_i)$ that is reached by a random walk that starts at $v$. We will use these probabilities to simulate a new random walk $\HX$ that we now  define.  For given trajectory $X(\omega)$ of walk $X$, $\HX(\omega)$ is equal to $\TX(\omega)$ (as defined before),  except that whenever $\TX(\omega)$ shortcuts some block visited by $X$, $\HX(\omega)$ contains only the first vertex visited in this block, as opposed to both first and last vertices retained in $\TX(\omega)$. $\HX$ is a shortcutting of $\TX$ and is thus a shortcutting of $X$ as well.

 It is not hard to see that by using  $Q_v(u)$ in a way completely analogous to the way we used $P_v(e)$ before, we can simulate the walk $\HX$ efficiently, and the expected length of this walk is bounded by the expected length of the walk $\TX$. However,  unlike $\TX$, $\HX$ does not necessarily posess all of the information needed to reconstruct the final arborescence. This shortcoming manifests itself whenever some $u$ is visited for the first time in $\HX$ directly after $\HX$ entered some $D_i$ after $\tau_i$ has already occurred. In this case, we know that the corresponding trajectory of the walk $X$ visited $u$ for the first time through some edge whose other end was in $D_i$ (and thus we should add it to our arborescence as $e_u$), but we don't know which one it was.

 To deal with this problem, we will define a stronger decomposition of $G$ whose properties will imply that
 the above failure to learn $e_u$ will occur only for a small number of vertices $u$.
  Then, at the end of our simulation of the walk $\HX$, we will employ a procedure that will compute the missing arcs in a manner that will not distort the desired distribution over output arborescences.

 We proceed now to formalizing the above outline.

 \subsection{Strong $(\phi, \gamma)$-decompositions}

The number of probabilities $Q_v(u)$ that we need to compute now depends on the number of vertices that are connected to the $D_i$s through some edge from $C$, rather than just the number of edges of $C$.  To control this, we introduce the following stronger graph decomposition:

\begin{definition}\label{def:decomp2}
A partition $(D_1,\ldots, D_k,S,C)$ of $G$ is a {\em strong ($\phi$, $\gamma$)-decomposition} of $G$ if:
\begin{enumerate}
\item $(D_1,\ldots, D_k,S,C)$ is a ($\phi$,$\gamma$)-decomposition,
\item   there are no edges between different $D_i$s (i.e. $S$ is a vertex multiway cut), and
\item  $|C(S)|\leq \phi |V(G)|$, where $C(S)$ is the set of vertices from $S$ that are connected by an edge to some $D_i$.
\end{enumerate}
\end{definition}

 We prove the following:

\begin{lemma}\label{lem:decomp2}
For any $G$ and any $\phi=o(1)$, there exists a strong $(\phi, \TO(1/\phi))$-decomposition of
$G$. Moreover, such a decomposition can be computed in time $\TO(m)$.
\end{lemma}

\begin{Proof}
We will use the ball-growing technique of Leighton and Rao \cite{LeightonRao} as presented by Trevisan \cite{Trevisan05approximationalgorithms}.
For a graph $H$, let $B_H(v,j)$ be the ball of radius $j$ around the vertex $v$ in $H$, i.e., let $B_H(v,j)$ consist of the subgraph of $H$ induced by all vertices in $H$ that are reachable by a path of length at most $j$ from $v$. Furthermore, let $R_H(v,j)$ be the set of vertices that are at distance exactly $j$ from $v$ in $H$. Finally, let $R_H^+(v,j)$ ($R_H^-(v,j)$ respectively) be the set $E(B_H(v,j+1))\setminus E(B_H(v,j))$ (the set $E(B_H(v,j))\setminus E(B_H(v,j-1))$ respectively).

Consider now the procedure presented in Table \ref{tab:partitioning algorithm}.
\begin{table}[!t]
\renewcommand{\arraystretch}{1.3}
\caption{A procedure finding a strong  $(\phi, \TO(1/\phi))$-decomposition of
$G$}
\label{tab:partitioning algorithm}
\centering
\fbox{
\begin{minipage}{6.25in}
\begin{itemize}
\item Set $H=G$, $S=\emptyset$, $C=\emptyset$, $D=\{\}$, $i=1$ and $t=\phi/(1-\phi)$
\item While $H\neq \emptyset$
\item \hspace{10pt} (* Ball-growing *)
\begin{itemize}
\item Choose an arbitrary $v\in H$, set $j=0$
\item $(*)$ As long as $|R_H(v,j+1)|>t|V(B_H(v,j))|$, $|R_H^+(v,j+1)|>t|E(B_H(v,j))|$ or $|R_H^-(v,j+1)|>t|E(B_H(v,j))|$:
\item \hspace{10pt} $j=j+1$
\item Let $j_i$ be the $j$ at which the above loop stops.  Add $R_H(v,j_i+1)$ to $S$, add all the edges incident to $R_H(v,j_i+1)$ (i.e. $R_H^+(v,j_i+1)\cup R_H^-(v,j_i+1)$) to $C$, and add $B_H(v,j_i)$ as component $D_i$ to $D$.
\end{itemize}
\item output the resulting partition $(D_1,\ldots ,D_k,S,C)$ of $G$
\end{itemize}
\end{minipage}
}
\end{table}
First, we note that this procedure can be implemented in nearly-linear time, since each edge is examined at most twice before it is removed from $H$. Moreover, an elementary charging argument shows that, in the resulting partition, $|C|\leq (1/(1+1/t)) |E(G)|=\phi |E(G)|$, and similarly $|C(S)|=|S|\leq \phi |V(G)|$. By construction, there are no edges between distinct $D_i$s. We want to argue now that for all $i$, $j_i\leq 3(1+\log |E(G)|/\log (1+t))$, which in turn would imply that all of the $D_i$s have diameter at most $6(1+\log |E(G)|)/\log (1+t)=6(1+\log |E(G)|)/\log (1/(1-\phi)))=O(\log m/(-\log (1-\phi)))=O(\log m/\phi)$, where we used Taylor expansion of $\log (1-x)$ around $x=0$ to get  this estimate.  To see why the above bound on $j_i$ holds, assume that it was not the case for some $i$ and $v$. Then, during the corresponding ball-growing procedure, a particular one of the three conditions from $(*)$ must have been triggered more than $j_i/3=1+\log |E(G)|/\log (1+t)$ times. If this condition was $|R_H(v,j+1)|>t|V(B_H(v,j))|$, then, since we never remove vertices from our ball that is being grown and $B_H(v,0)$ has one vertex, the final ball $B_H(v,j_i)$ has to have at least $(1+t)^{j_i/3}>|E(G)|\geq |V(G)|$ vertices, which is a contradiction. Similarly, if $|R_H^+(v,j+1)|>t|E(B_H(v,j))|$ ($|R_H^-(v,j+1)|>t|E(B_H(v,j))|$ respectively) was the condition in question, then $|E(B_H(j_i,v))|>(1+t)^{j_i/3}\geq |E(G)|$, which is a contradiction as well. Thus we may conclude that the above bound on $j_i$ holds, and all $D_i$s have diameter $\TO(1/\phi)$, as desired.

At this point, we know that the partition of $G$ that we obtained satisfies all of the properties of a strong $(\phi,\TO(1/\phi))$-decomposition except possibly the one that asserts that there is no $D_i$ such that $|E(D_i)|$ is smaller than the number $|C(D_i)|$ of edges from $C$ incident to $D_i$. However, if such $D_i$ exist, then we can just add them to our cut, i.e., we add $V(D_i)$ to $S$, and $E(D_i)$ to $C$. Note that the size of $C$ can at most triple as a result of this operation, since an edge that is initially in $C$ can be incident to at most two of the $D_i$, and edges $E(D_i)$ are by definition not incident to any other $D_j$. Similarly, $C(S)$ does not increase as a result of adding $V(D_i)$ to $S$. We may therefore conclude that the decomposition returned by this algorithm is indeed a strong $(\phi, \TO(1/\phi))$-decomposition of $G$.

\end{Proof}

From now on we fix some {\em strong} $(\phi, \gamma)$-decomposition of $G$.

\subsection{Computing $Q_v(u)$}

 By using an approach similar to the one that we used when computing the $P_v(e)$ values, we get the following lemma.

\begin{lemma}\label{lem:probabilities2}
Given a strong $(\phi,\gamma)$-decomposition of $G$, we can compute multiplicative $(1+\varepsilon)$-approximations of all $Q_v(u)$ in time $\TO(\phi
mn \log 1/\varepsilon)$.
\end{lemma}

\begin{Proof}
Let us fix some $D=D_i$, let $S'$ be the set of vertices in $w\in S$ with $|C_i(w)|>0$, and let us fix some $u\in S'$. Consider now a graph $D'$ that we obtain from $D\cup S'$  by merging all vertices in $S'\setminus \{u\}$ into one vertex called $u^*$.  For any given vertex $v\in D$, $Q_v(u)$ is exactly the probability that a random walk in $D'$ started at
$v$ will hit $u$ before it hits $u^*$. We can quickly compute such probabilities using
electrical flows.

More precisely, (see e.g. \cite{Lov_randomwalks}) if we treat $D'$ as electrical circuit in which
we impose voltage of $1$ at $u$ and $0$ at $u^*$, then the voltage achieved at $v$ in such
electrical flow is equal to $Q_v(e)$. We can compute  a $(1+\varepsilon)$-approximation of such
a flow in time $\TO(|E(D')|\log 1/\varepsilon)$ using the linear system solver of Spielman and  Teng~\cite{SpielmanTeng_solving}. To do so, let $L$ be the Laplacian of $D'$, where we let the first two rows correspond to $u$ and $u^*$, respectively. Furthermore, let  $\mathbf{i}_{ext}\in \mathbb{R}^{|V(D')|}$ be the vector that is $1$ in its first coordinate, $-1$ in its second coordinate, and zero everywhere else. Let $\mathbf{v'}$ be the solution to the linear system $L \mathbf{v'} =\mathbf{i}_{ext}$, which we can approximately find in nearly-linear time using the solver from~\cite{SpielmanTeng_solving}.    We obtain our desired vector of voltages by subtracting $v'_2$ from all if its coordinates and dividing them by $v_{1}'-v_{2}'$.

Our algorithm computes such voltages for each $u\in S$ and all $D_i$ with $|C_i(u)|>0$,  and from
each such computation it stores the probabilities $Q_v(u)$ for all vertices $v\in V(D_i)$ that we are interested in.
The running time of such procedure is bounded by $\TO(|C(S)|\sum_i |E(D_i)|\log
1/\varepsilon)=\TO(\phi m n\log 1/\varepsilon)$, as desired.
\end{Proof}

\subsection{Coping with shortcomings of $\HX$}

As mentioned above, the walk $\HX$ can be simulated more efficiently than the walk $\TX$, but it does not have all the information needed to construct the arborescence that would be generated by the walk $X$ that $\HX$ is meant to shortcut. This lack of information occurs only when some vertex $u$ is visited for the first time by $\HX$ immediately after $\HX$ visited a component $D_i$ after time $\tau_i$ has already occurred.  Note that, by the properties of the strong $(\phi,\gamma)$-decomposition that we are using,  it must be the case that $u\in C(S)$ and $|C(S)|\leq \phi n$. This shows that $\HX$ fails to estimate the arcs $e_u$ for a small fraction of vertices of $G$. We prove now that, in this case, these missing arcs can be reconstructed efficently and in a way that preserves the desired distribution over arborescences.

\begin{lemma}\label{lem:reconstructing}
For a trajectory $\HX(\omega)$ that starts at vertex $s$, let $F(\HX(\omega))$ be the set of arcs $e_v$, for $v\notin C(S)\cup \{s\}$, corresponding to this trajectory.  Let $F(\HX(\omega))^*$ be the set of all arborescences $H$ rooted at $s$ such that $e_H(v)=e_v$ for $v\notin C(S)\cup \{s\}$. Then, given $F(\HX(\omega))$, we can generate a random arborescence from $F(\HX(\omega))^*$ in time $O(m+(\phi n)^{2.376})$.
\end{lemma}

\begin{Proof}
For brevity, let us define $F:=F(\HX(\omega))$. Now, let $H_1,\ldots ,H_r$, with $s\in H_1$, be the decomposition of $F$ into weakly connected components. We construct a {\em directed} graph $G(F,s)$ as follows. $G(F,s)$ has a vertex $h_j$ for each $H_j$. $E(G(F,s))$ is the set of all arcs $(h_j,h_l)$ such that there exists $v,u\in G$ with $v\in H_j$, $u\in H_l$ and $u\in C(S)\setminus\{s\}$. By definition,  if $H'$ is an arborescence in $G(F,s)$ rooted at $h_1$, then $H'\cup F$ is an arborescence in $G$ rooted at $s$ and $H'\cup F\in F^*$. Moreover, for any arborescence $H\in F^*$, $H'=\{e_H(v)\  | \ v \in C(S)\setminus \{s\}\}$ is an arborescence in $G(F,s)$ rooted at $h_1$. So, if we use the algorithm of Colbourn {\em et al.} \cite{Colbourn} to generate random arborescence $H'$ in $G(F,s)$ rooted at $h_1$, then $H'\cup F$ is a random arborescence from $F^*$. Since $|V(G(F,s))|=|C(S)|\leq \phi n$ and the algorithm from \cite{Colbourn} works in time $O(|V(G(F,s))|^{2.376})$, the lemma follows.

\end{Proof}

\subsection{Proof of Theorem \ref{thm:main} }

By the connection explained in section \ref{sec:cond_rand}, it is sufficient  to devise a procedure that generates $\delta$-random arborescences. We do this as follows. We fix $\phi=1/\sqrt{n}$, and, using Lemma \ref{lem:decomp2}, we get a strong $(1/\sqrt{n},\TO(\sqrt{n}))$-decomposition of $G$ in $\TO(m)$ time. Now, using Lemma \ref{lem:probabilities2}, we compute $\varepsilon$-approximations of all  of the probabilities $Q_v(u)$ in time $\TO(m\sqrt{n} \log 1/\varepsilon)=\TO(m\sqrt{n} \log 1/\delta)$, where we set $\varepsilon=O(\delta/n^{5})$. At this point, we can use completely analogous reasoning to that used in Lemmas \ref{lem:wrapup1} and Lemma \ref{conj} to prove that  we can simulate ($\delta$-approximately) the walk $\HX$ in expected time $\TO(m\sqrt{n})$ (where we use in particular the fact that $\HX$ is a shortcutting of $\TX$). Having done this, we look at the directed forest $F$ consisting of arcs $e_v$ for $v\notin C(S)\cup \{s\}$ as defined by our simulation of $\HX$. We then use  the procedure from Lemma \ref{lem:reconstructing} to get an arborescence $T$ of $G$ in time $O(m+n^{2.376/2})=O(m\sqrt{n})$. The whole algorithm therefore runs in expected time $\TO(m\sqrt{n} \log 1/\delta)$.

 To see that the above algorithm generates a (conditionally) $\delta$-random arborescence, let us consider some arborescence $T$ of $G$ rooted at some vertex $s$ and condition what will follow on the event that the algorithm outputs an arborescence rooted at $s$ . Let $F(T)=\{e_T(v)\ | \  v\notin C(S)\cup\{s\}\}$. By the fact that $\HX$ is a ($\delta$-approximate) shortcutting of the walk $X$ and by Theorem \ref{thm:randwalkalg}, we know that $F(\HX(\omega))=F(T)$ for  at least a $(1-\delta) |{F(T)^*}|/|\mathcal{T}_s(G)|$ and at most a $(1+\delta) |{F(T)^*}|/|\mathcal{T}_s(G)|$ fraction of trajectories, where $F(T)^*$ is the set of all arborescences compatible with $F(T)$ in the sense of Lemma \ref{lem:reconstructing}. Since the procedure from Lemma \ref{lem:reconstructing} generates a random arborescence from ${F(T)^*}$, $T$ is generated with probability  at least $(1-\delta) /|\mathcal{T}_s(G)|$ and at most $(1+\delta) /|\mathcal{T}_s(G)|$. This concludes the proof of Theorem \ref{thm:main}.